\magnification=\magstep1
\baselineskip=20pt
\centerline{Implications of the Visible and X-Ray Counterparts to GRB970228}
\bigskip
\centerline{J. I. Katz\footnote*{Racah Institute of Physics, Hebrew
University, Jerusalem 91904, Israel}\footnote\dag{Department of Physics and
McDonnell Center for the Space Sciences, Washington University, St. Louis,
Mo. 63130}, T. Piran$^{\rm *}$, R. Sari$^{\rm *}$}
\bigskip
The gamma-ray burst source GRB970228$^{1,2}$ has been observed after a
delay of 8--12 hours in X-rays$^3$ and after one day in visible and near
infrared light$^4$.  This marks the first detection of emission at lower
frequencies following the gamma-ray observation of a GRB and the first
detection of any visible counterpart to a GRB.  We consider possible
delayed visible and X-ray emission mechanisms, and conclude that the
intrinsic gamma-ray activity continued at a much reduced intensity for at 
least a day.  There are hints of such continued activity in other GRB, and
future observations can decide if this is true of GRB in general.
The observed multi-band spectrum of GRB970228 agrees with the predictions of
relativistic shock theory when the flux is integrated over a time longer 
than that required for a radiating electron to lose its energy.

Several mechanisms for the continuing X-ray emission of GRB970228 should be
considered.  The relativistic particles required to explain a GRB will
collide with a surrounding dilute medium, a process which has been 
suggested$^5$ as the source of the gamma-ray emission itself.  Simple
analytic models of such external shocks which follow the degradation of the 
energy of the particles and fields lead to a predicted time scale of 
emission $\propto \nu_{obs}^{-5/12}$, where $\nu_{obs}$ is the frequency of
observation$^6$.  Similarly, a simple model$^7$ in which relativistic
electrons radiate their energy in a constant magnetic field predicts a time 
scale $\propto \nu_{obs}^{-1/2}$.  These classes of models may be excluded
because the observed duration of X-ray emission$^3$ is roughly 1000 times 
the reported gamma-ray duration$^1$, despite a ratio in $\nu_{obs}$ of only
$\sim 50$.  Another alternative model of the X-ray emission, thermal 
bremsstrahlung (as in a supernova remnant), may also be excluded because
the required power $\sim 10^{45}$ erg/s (at cosmological redshift $z = 
0.498^{8,9}$) would require an unachievable particle density $> 10^{10}$
cm$^{-3}$ even if the maximum plausible mass of 1 $M_\odot$ is radiating.

Instead, we suggest that the observed brief intense gamma-ray emission of
a GRB is only the tip of an iceberg; it emits gamma-rays at a much lower 
level for time of order a day following (and perhaps preceding) the intense 
outburst.  GRB detectors necessarily have high backgrounds because they must
have very broad angular acceptance; these high backgrounds, lack of angular 
discrimination and necessarily short integration times imply high thresholds
for detected flux, making the continuing weak gamma-ray emission difficult
to observe by the burst detector.  The X-ray and visible radiation is then
produced by the same mechanism as the gamma rays, simultaneously with their
continuing emission.  

There is independent evidence for continuing gamma-ray activity in GRB,
with durations longer than the usual values $< 1000$ s$^{10}$.
GRB940217 was observed$^{11}$ to emit energetic gamma-rays over a duration 
of $\approx 5000$ seconds.  The group of four GRB observed$^{12,13}$ October
27--29, 1996, apparently from a common source, may equally well be described
as repeating bursts or as a single burst lasting two days with brief periods
of high intensity amidst a much longer period of undetectably low intensity.
The occasional observation of ``precursors'' some time before the peak 
emission of a GRB$^{14}$ may also indicate a longer period of weak activity.

The hypothesis that many GRB last a day or more is consistent with the
demon-stration$^{15}$ that the observed complex time structure of GRB on
scales of seconds must be attributed to variations in the power of their
energy source.  Without a better understanding of this central engine,
durations of a day are no less plausible than durations of tens of seconds.

Our suggestion that the X-ray and visible emission of GRB970228 was the
consequence of continuing weak gamma-ray activity predicts that this
gamma-ray emission should be observable in similar bursts by suitable 
instruments.  It also leads to a specific prediction for its spectrum, which
may be roughly tested with the data at hand.  The instantaneous spectrum of 
a relativistic shock is predicted$^{16,17}$ to be $F_\nu \propto \nu^{1/3}$.
The spectrum integrated over the radiative decay of the electrons' energy is
predicted$^{18}$ to be $F_\nu \propto \nu^{-1/2}$.  Observations during the
brief phases of GRB during which BATSE obtained data have shown soft
gamma-ray spectra between these limits, with $F_\nu \propto \nu^{-1/2}$
found when the data are integrated over many subpulses, allowing time for
shock-heated electrons to radiate their energy$^{18}$.  This prediction
should be applicable to X-ray$^3$ and visible$^4$ data obtained over much
longer periods of integration.

The data$^{2-4}$ from GRB970228 are collected in the Figure.  The V and I
band data were obtained$^4$ about one day after the observed GRB, and the
X-ray data$^3$ were integrated over the period 8--12 hours after the GRB.
Because these delays are similar, these data should be comparable.  The
solid line shows the predicted $F_\nu \propto \nu^{-1/2}$ slope, fitted
(arbitrarily) to the V data point.  The V, I, and X-ray data are all
consistent with the predicted slope, confirming the hypothesis.

The only available gamma-ray data is that obtained$^{1,2}$ in the most
intense 80 s of the GRB.  This is not directly comparable to the visible,
near infrared and X-ray fluxes obtained after longer delays, but it is
possible to define an equivalent mean gamma-ray flux by dividing the
gamma-ray fluence$^2$ by the estimated one day X-ray and visible decay time.
This has been included in the Figure, and is also consistent with the
predicted slope.  The consistency of this equivalent mean gamma-ray flux with
the extrapolated spectrum leads to the prediction that the actual mean flux 
is comparable to this value.  It may be too late to test this prediction for
GRB970228, but a similar analysis may be performed on future GRB.

Acknowledgments: We thank NASA, NSF and the US-Israel BSF for support.
JIK thanks Washington University for the grant of sabbatical leave and the
Hebrew University for a Forchheimer fellowship.
\vfil
\eject
\centerline{References}
\bigskip
\item{1.}Costa, E., {\it et al.} {\it IAU Circ.} 6572 (1997). 
\item{2.}Palmer, D., {\it et al.} {\it IAU Circ.} 6577 (1997).
\item{3.}Costa, E., {\it et al.} {\it IAU Circ.} 6576 (1997).
\item{4.}Groot, P. J., {\it et al.} {\it IAU Circ.} 6584 (1997).
\item{5.}Rees, M. J. \& M\'esz\'aros, P. {\it MNRAS} {\bf 258}, 41p--43p 
(1992). 
\item{6.}Katz, J. I. {\it Astrophys. J.} {\bf 422}, 248--259 (1994).
\item{7.}Sari, R., Narayan, R. \& Piran, T. {\it Astrophys. J.} {\bf 473},
204--218 (1996).
\item{8.}Wagner, R. M., Foltz, C. B. \& Hewett, P. {\it IAU Circ.} 6581 
(1997).
\item{9.}Metzger, M. R., Kulkarni, S. R., Djorgovski, S. G., Gal, R., 
Steidel, C.  C. \& Frail, D. A. {\it IAU Circ.} 6582 (1997).
\item{10.}Fishman, G. J., {\it et al.} {\it Astrophys. J. Suppl.} {\bf 92}, 
229--283 (1994).
\item{11.}Hurley, K., {\it et al.} {\it Nature} {\bf 372}, 652--654 (1994).
\item{12.}Meegan, C., {\it et al.} {\it IAU Circ.} 6518 (1996).
\item{13.}Connaughton, V., {\it et al.} {\it Proc. 18th Texas Symposium on
Relativistic Astrophysics} (N. Y. Acad. Sci.: New York) in press (1997).
\item{14.}Koshut, M. {\it et al.} {\it Astrophys. J.} {\bf 452}, 145--155 
(1995).
\item{15.}Sari, R. \& Piran, T. {\it Astrophys. J.} in press 
(astro-ph/9701002) (1997).
\item{16.}Katz, J. I. {\it Astrophys. J. Lett.} {\bf 432}, L107--L109 (1994).
\item{17.}Tavani, M. {\it Astrophys. Sp. Sci.} {\bf 231}, 181--185 (1995).
\item{18.}Cohen, E., Katz, J. I., Piran, T., Sari, R., Preece, R. D. \& 
Band, D. L. {\it Astrophys. J.} submitted (astro-ph/9703120) (1997).
\vfil
\eject
Figure Caption: Fluxes of GRB970228 in soft gamma-ray$^2$, X-ray$^3$, V$^4$
and I$^4$ bands.  The soft gamma-ray flux is actually the fluence 
measured$^2$ in a 3.6 s spike in intensity, divided by the one day decay
time roughly characteristic of the X-ray and visible fluxes.  The straight
line has the predicted $-1/2$ slope, normalized to the V band data.
\vfil
\eject
\bye
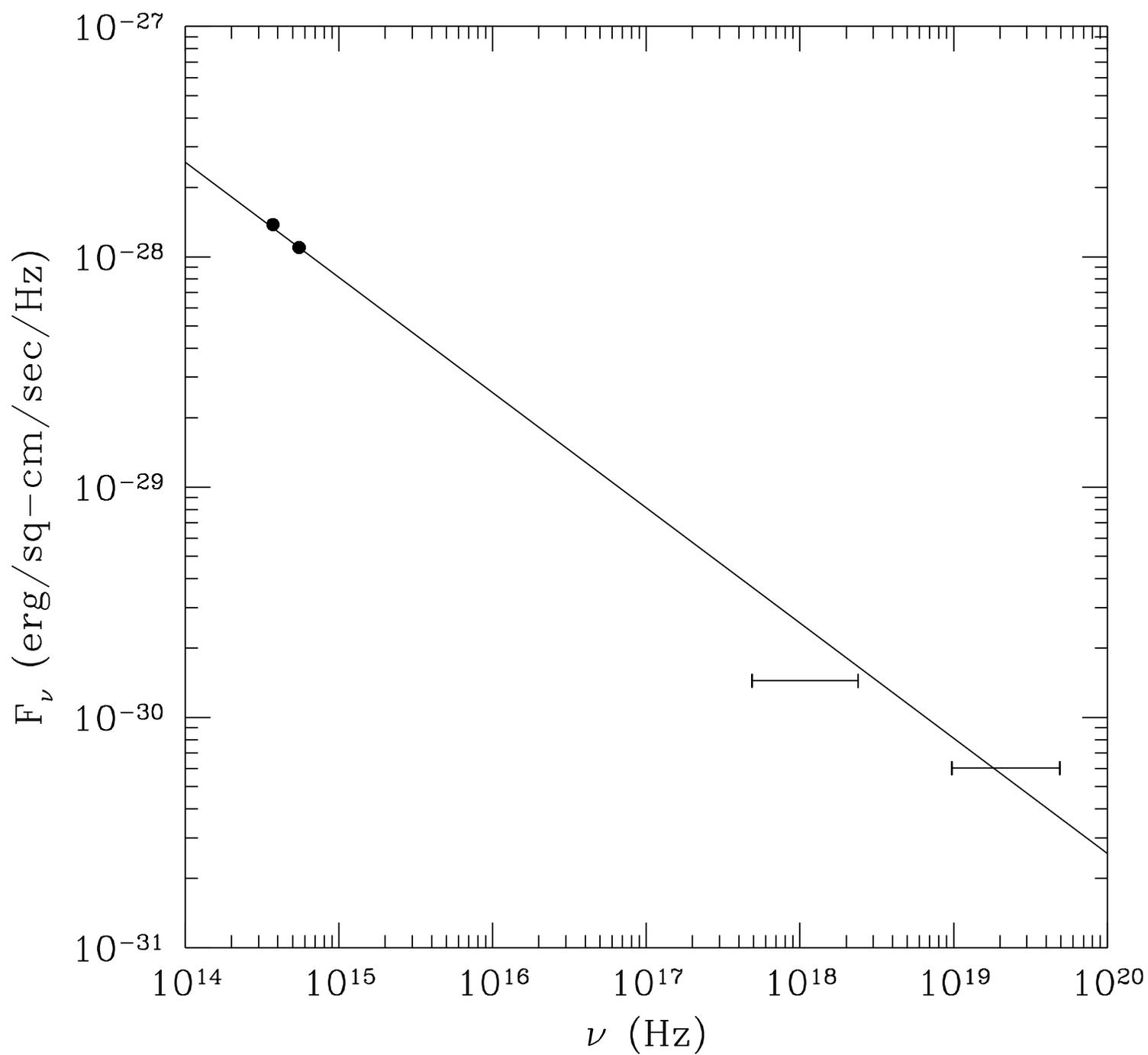
\end